\documentclass[aps,prb,reprint,groupedaddress,footinbib,citeautoscript]{revtex4-1}

\usepackage{graphicx}
\usepackage{dcolumn}
\usepackage{bm}
\usepackage{bbm}
\usepackage{amsfonts}

\usepackage{amsmath}

\newcommand* {\ket}[1]{\ensuremath{| {#1} \rangle}}
\newcommand\redout{\bgroup\markoverwith{\textcolor{red}{\rule[.5ex]{2pt}{0.4pt}}}\ULon}

\usepackage[colorlinks=true,citecolor=blue]{hyperref}
\hypersetup{colorlinks=true,citecolor=blue,linkcolor=red,urlcolor=blue}

\graphicspath{{.}{./Figures/}}

\begin{document}

\title{Unconventional superconductivity from magnetism in transition metal
dichalcogenides}

\author{M.~A. Rahimi}
\affiliation{Department of Physics, Institute for Advanced Studies in Basic Sciences
(IASBS), Zanjan 45137-66731, Iran}

\author{A.~G. Moghaddam}
\email{agorbanz@iasbs.ac.ir}
\affiliation{Department of Physics, Institute for Advanced Studies in Basic Sciences
(IASBS), Zanjan 45137-66731, Iran}

\author{C. Dykstra}
\affiliation{School of Chemical and Physical Sciences and MacDiarmid Institute
for Advanced Materials and Nanotechnology, Victoria University of Wellington,
PO Box 600, Wellington 6140, New Zealand}

\author{M. Governale}
\email{michele.governale@vuw.ac.nz}
\affiliation{School of Chemical and Physical Sciences and MacDiarmid Institute
for Advanced Materials and Nanotechnology, Victoria University of Wellington,
PO Box 600, Wellington 6140, New Zealand}

\author{U. Z\"ulicke}
\email{uli.zuelicke@vuw.ac.nz}
\affiliation{School of Chemical and Physical Sciences and MacDiarmid Institute
for Advanced Materials and Nanotechnology, Victoria University of Wellington,
PO Box 600, Wellington 6140, New Zealand}

\date{\today}

\begin{abstract}

We investigate proximity-induced superconductivity in monolayers of transition metal 
dichalcogenides (TMDs) in the presence of an externally generated exchange field. A 
variety of superconducting order parameters is found to emerge from the interplay of
magnetism and superconductivity, covering the entire spectrum of possibilities to be
symmetric or antisymmetric with respect to the valley and spin degrees of freedom,
as well as even or odd in frequency. More specifically, when a conventional
\emph{s}-wave superconductor with singlet Cooper pairs is tunnel-coupled to the
TMD layer, both spin-singlet and triplet pairings between electrons from the same
and opposite valleys arise due to the combined effects of intrinsic spin-orbit
coupling and a magnetic-substrate-induced exchange field. As a key finding, we
reveal the existence of an exotic even-frequency triplet pairing between equal-spin
electrons from different valleys, which arises whenever the spin orientations in
the two valleys are noncollinear. All types of superconducting order turn out to be
highly tunable via straightforward manipulation of the external exchange field.

\end{abstract}

\maketitle

\section{Introduction}

Understanding possible mechanisms for the coexistence and interplay of superconductivity 
with magnetism has been one of the most long-standing and intensely studied questions
in condensed-matter physics. Conventional \textit{s}-wave BCS superconductivity is known
to be fragile against magnetic perturbations~\cite{anderson59,kulic85} due to its 
spin-singlet character. Instances where \textit{s}-wave superconductivity coexists
with magnetism typically require spatially inhomogeneity. Pertinent examples are the
theoretically proposed Fulde-Ferell-Larkin-Ovchinikov (FFLO) superconducting
state~\cite{ff64,lo65,fflo-rev} that oscillates in space, or hybrid structures of
superconducting and magnetic materials~\cite{buzdin}. Alternatively, unconventional
\textit{p}-wave superconductivity~\cite{Mineev1999} can accommodate magnetism because
it enables Cooper-pairing of electrons with the same spin. Originally developed to
understand superfluidity in $^3$He~\cite{richardson,berezinskii,leggett75,leggett-book},
and later surmised to describe superconducting order in Sr$_2$RuO$_4$~\cite{mackenzie03},
triplet pairing has recently attracted particular attention in the context of proximity 
effects in superconductor-ferromagnet heterostructures\cite{buzdin,efetov,klapwijk06}.
Especially intriguing features include the emergence of odd-frequency triplet pairing
due to the interplay of superconducting correlations with inhomogeneous non-collinear
magnetizations~\cite{efetov,buzdin,klapwijk06,kadigrobov01,bergeret01,volkov03,birdge10,
robinson10}, and the possibility for exotic Majorana bound states to be hosted in
systems where the three ingredients of spin-orbit-coupling, Zeeman splitting, and 
superconducting correlations are present~\cite{das-sarma10,oreg10,alicea12,
kouwenhoven12,beenakker13}.

On the other hand, the experimental realization of atomically thin crystalline
materials has opened up an entirely new era of mesoscopic and nanoscale 
physics~\cite{geim13,xu13,butler13}. This development started with the isolation of
graphene~\cite{novo04,novo05nat,kim05,geim07,neto09} and rapidly continued with
subsequent discoveries of related structures including silicene~\cite{sahin09,vogt12}
and transition metal dichalcogenide monolayers (TMD-MLs)~\cite{mak10,chhowalla} such
as monolayer MoS$_2$. TMD-MLs have been the focus of great interest due to the
intriguing effects associated with the coupling of spin and valley degrees of 
freedom~\cite{xu14}. Their very large intrinsic spin-orbit coupling (SOC) suggest these
materials can be used as a building blocks for spintronic applications~\cite{zhu11,
yuan13,riley14,burkard14prx,roldan14-2d,loss13}. Furthermore, a significant valley
polarization can be induced by circularly polarized light~\cite{zeng12,mak12,behnia12, 
cao12}, leading to the observation of the valley Hall effect and growing possibilities
for various valleytronics and optoelectronics applications~\cite{mceuen14,niu07,xiao12}.
Most recently, both experimental and theoretical investigations have revealed that 
monolayer and few layer TMDs exhibit superconducting signatures under certain 
circumstances. In particular, ionic-gated MoS$_2$, MoSe$_2$, and WS$_2$ undergo a 
superconducting phase transition at intermediate dopings~\cite{super-tmd-1,
super-tmd-2,super-tmd-3}. Recent observations also revealed an unusual superconducting 
behavior in few-layer NbSe$_2$ and MoS$_2$ that is robust against very large in-plane 
magnetic fields~\cite{ising-1,ising-3,ising-2}. This so-called Ising superconductivity 
originates from the intrinsic SOC and the two-dimensional nature of these
materials~\cite{jarillo-herrero,law16}.

Previous theoretical studies have focused on the proximity effect, topological intrinsic
superconductivity, and the Josephson effect in TMD-ML~\cite{law16,liu13,guinea13,yuan14,
rameshti14,rossi16}. In the present work, we provide a complete understanding of the
interplay of superconducting proximity effects with magnetism in TMD. Our model system
consists of a TMD-ML that is coupled vertically to a superconducting material and also
subject to a finite spin exchange induced by a ferromagnetic substrate or external 
magnetic field. The interplay of SOC, the two-valley band structure, and the externally 
induced spin splitting results in exotic proximity effects that are different from
common superconductor-ferromagnet heterostructures. In particular, we show that 
spin-triplet pairings that are odd with respect to either frequency or the valley
degree of freedom are induced without any inhomogeneous and noncollinear magnetizations.
We also identify the existence of induced intra-valley pairings, which are particular
instances of the generic pair-density-wave order associated with Cooper pairing at
finite momentum~\cite{fflo-rev} that has been discussed previously, e.g., for
high-$T_\mathrm{c}$ superconductors~\cite{fradkin15,lee14}, other 2D
materials~\cite{roy10,zhou13,tsu16}, and Weyl semimetals~\cite{cho12}.

\begin{figure}[t]
\includegraphics[width=0.95\columnwidth]{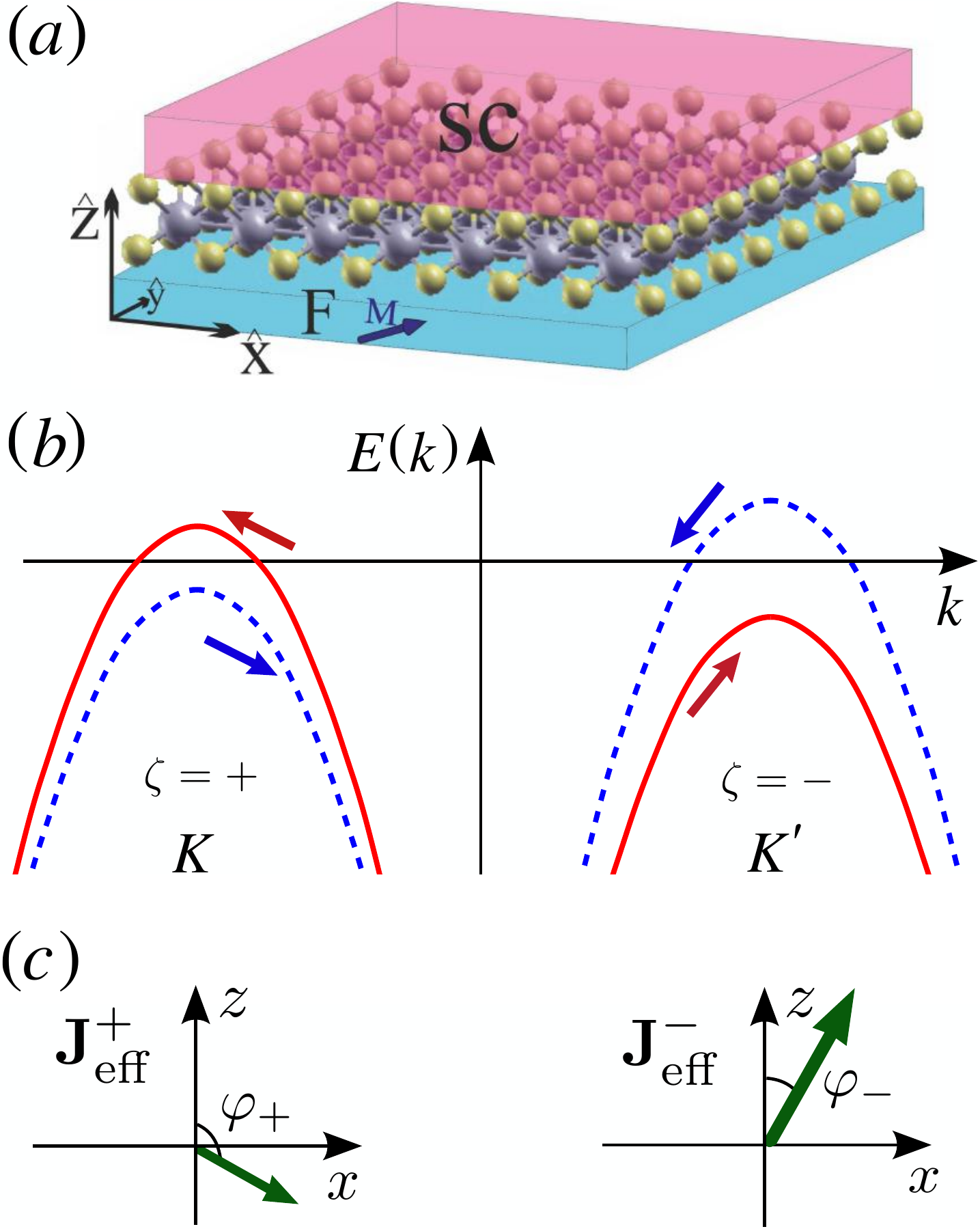}
\caption{\label{fig1}%
Structure and properties of our system of interest. (a)~Schematics of the hybrid
superconductor--TMD-ML--magnetic-insulator system. (b)~Valence-band structure of the
TMD-ML in the presence of an externally induced exchange field ${\bf J}$. The spin
states of each subband are indicated for an exchange field of the form ${\bf J}=(J_x,
0,J_z)$. (c)~The effective exchange fields $J_{\rm eff}^{\zeta}$ in the two valleys
arising from the combination of ${\bf J}$ and SOC-induced spin splitting, which point 
at angles $\varphi_{\pm}=\arctan[J_x/( J_z\mp\lambda)]$ with respect to the $z$ axis.}
\end{figure}

\section{Basic theory and formalism}

We consider a TMD-ML on a ferromagnetic substrate that is tunnel-coupled to a
conventional superconductor as sketched schematically in Fig.~\ref{fig1}(a). The effect
of the ferromagnetic substrate is to induce an exchange field ${\mathbf J}$ inside the
TMD parallel to the substrate magnetization ${\mathbf M}$. We focus on the case of a
hole-doped TMD-ML and therefore only consider this material's two spin-split valence
bands at each valley. The relevant band dispersions for this case are shown in
Fig.~\ref{fig1}(b). The low-energy Hamiltonian of this system in the vicinity of the
${\mathbf K}$ and $\mathbf{K'}$ points has the form
\begin{subequations}
\begin{eqnarray}
&& {\cal H}_{\rm TMD}=\sum_{\bf k,\zeta}  \hat{\bm \phi}^\dagger_{{\bf k},\zeta}
~[\hat{h}_{\zeta}({\bf k})\otimes \hat{\tau}_z]~
 \hat{\bm \phi}_{{\bf k},\zeta} \quad ,\\
 \label{htmd}
&& \hat{h}_{\zeta}({\bf k})=E_{\bf k} \hat{\sigma}_0
-{\bf J}\cdot \hat{\bm \sigma} +\zeta \lambda \hat{\sigma}_z \quad ,\\
&& E_{\bf k}= -\frac{\hbar ^2 |{\bf k}|^2}{2m_\mathrm{v}}+E_{\text{v}} \quad .
\end{eqnarray}
\end{subequations}
The multi-component particle operator $\hat{\bm \phi}^{\dagger}_{{\bf k},\zeta}=
\left( c^{\dagger}_{{\bf k}\zeta\uparrow}, c^{\dagger}_{{\bf k}\zeta\downarrow},
c_{-{\bf k}\zeta\uparrow}, c_{-{\bf k}\zeta\downarrow}\right)$ is formed from the
creation and annihilation operators $c^{\dagger}_{{\bf k}\zeta s}$ and $c_{{\bf k}
\zeta s}$ for electron states with wave vector ${\bf k}$, spin $s=\,\,\uparrow,
\downarrow$, and valley index $\zeta=\pm 1$ distinguishing the $\mathbf{K}$ and
$\mathbf{K'}$ points, respectively. The parameters $m_\mathrm{v}$ and $\lambda$
correspond to the valence-band effective mass and the strength of intrinsic spin-orbit
coupling (SOC). The energy $E_{\text{v}}$ measures the distance of the valence-band
edge for vanishing $\mathbf{J}$ and $\lambda$ from the chemical potential of the
superconductor. To be specific, we set $E_{\text{v}}=0$ in the following, which implies
that the chemical potential is in the middle between the spin-split valence bands. See
Fig.~\ref{fig1}(b). The Pauli matrices $\tau_i$ and $\sigma_i$ ($i=x,y,z$) operate on
the Nambu and spin spaces, respectively. Corresponding identity matrices are denoted by
$\tau_0$ and $\sigma_0$. The combination of the valley-antisymmetric SOC and the
valley-symmetric exchange field leads to two different effective spin splittings in the
$\mathbf{K}$ and $\mathbf{K'}$ valleys,
\begin{align}
\label{eq:jeff}
\mathbf{J}^{\zeta}_{\rm eff}=\mathbf{J}-\zeta\lambda\hat{\bf z} \quad .
\end{align}
This is illustrated in Fig.~\ref{fig1}(c). 

The electronic degrees of freedom of a conventional \textit{s}-wave superconductor are
described by the mean-field BCS Hamiltonian
\begin{equation}
\label{eq:bcs}
\mathcal{H}_{\rm SC}=\sum_{\bf q} \hat{\bm \psi}^{\dagger}_{\bf q}\left( \xi_{\bf q} 
\hat{\sigma}_0 \otimes \hat{\tau_z}  - \Delta_{\rm SC}  \hat{\sigma}_y \otimes 
\hat{\tau}_y \right) \hat{\bm \psi}^{\dagger}_{\bf q} \quad,
\end{equation}
where the particle operator $\hat{\bm \psi}^{\dagger}_{\bf q}=\left( 
d^{\dagger}_{{\bf q}\uparrow}, d^{\dagger}_{{\bf q}\downarrow}, d_{-{\bf q}\uparrow},
d_{-{\bf q}\downarrow} \right)$ is formed from creation and annihilation operators for
electron states inside the superconductor. The gap and single-particle dispersion
relation of the superconductor are denoted by $\Delta_\mathrm{SC}$ and $\xi_\mathbf{q} 
=\hbar^2 q^2/(2m_{\text{S}})-E_{\text{FS}}$, with $E_{\text{FS}}$ being the Fermi
energy of electrons in the superconductor. The tunnel coupling between the TMD-ML and
the superconductor is modelled by the tunnelling Hamiltonian
\begin{subequations}
\begin{eqnarray}\label{thamiltonian}
\mathcal{H}_T&=&\sum_{{\bf q},{\bf k}} 
(\hat{\phi}^{\dagger}_{{\bf k},+},\hat{\phi}^{\dagger}_{{\bf k},-}) 
T_{\mathbf{k},\mathbf{q}} \hat{\psi}_{\bf q}  +{\rm h.c.} \quad , \\[0.1cm]
T_{\mathbf{k},\mathbf{q}}&=&\begin{pmatrix}
t_{\mathbf{k},\mathbf{q},+}\hat{\sigma}_0\otimes \hat{\tau}_z \\
t_{\mathbf{k},\mathbf{q},-} \hat{\sigma}_0\otimes \hat{\tau}_z
\end{pmatrix} \quad ,
\end{eqnarray}
\end{subequations}
where $T_{\mathbf{k},\mathbf{q}}$ is an $8\times4$ matrix that couples the electronic
states from the two materials. We assume generally different tunnel couplings
$t_{\mathbf{k},\mathbf{q},+}$ and $t_{\mathbf{k},\mathbf{q},-}$ for the two valleys
$\mathbf{K}$ and $\mathbf{K'}$, respectively. We will furthermore focus on the case of
constant momentum-independent tunnel couplings $t_{\mathbf{k},\mathbf{q},\zeta} \equiv
t_{\zeta}$. 

In order to investigate the proximity-induced superconducting correlations in the
TMD-ML arising from its tunnel coupling to the superconductor, we make use of the
Matsubara Green's-function formalism.\cite{bruusbook} In the absence of tunneling, the
Green's function of the superconductor is given by 
\begin{eqnarray}
&&G^{0}_{\rm SC}({\bf q},i\omega_n)=
\begin{pmatrix}
 \hat{g}({\bf q},i\omega_n) &  \hat{f}({\bf q},i\omega_n)  \\
\hat{f}^\dag({\bf q},i\omega_n) & -\hat{g} (-{\bf q},-i\omega_n)
\end{pmatrix},\\[0.1cm]
&&\hat{g}({\bf q},i\omega_n)=\frac{(i\omega_n+\xi_{\bf q})\hat{\sigma}_0 }{(i\omega_n)^2 
-\varepsilon_{\bf q}^2}~,~~~~ \hat{f}({\bf q},\omega)=\frac{i\Delta_{\rm SC}
\hat{\sigma}_y}{(i\omega_n)^2 -\varepsilon_{\bf q}^2}, \nonumber
\end{eqnarray}
where $\varepsilon_\mathbf{q}=\sqrt{\xi_{\bf q}^2+\Delta_{\rm SC}^2}$ is the
quasiparticle excitation spectrum. The bare Green's function $G^{0}_{\rm TMD}$ of the
isolated TMD consist of four diagonal blocks, each having a certain valley index and
corresponding to either electrons or holes,
\begin{eqnarray}
G^{0}_{\mathrm{TMD}}=
\begin{pmatrix}
\hat{g}^{(+)}_e&0&0&0\\
0&\hat{g}^{(+)}_h&0&0\\
0&0&\hat{g}_e^{(-)}&0\\
0&0&0&\hat{g}_h^{(-)}\\
\end{pmatrix}.
\end{eqnarray}
The blocks on the diagonal are $2\times 2$ matrices in spin space, given by the
expressions $\hat{g}^{(\zeta)}_{e,h}({\bf k},i\omega_n)=[i\omega_n\pm h_{\zeta}
(\bf {k})]^{-1}$. 

The full Green's function ${\mathcal G}_{\text{TMD}}$ of the TMD-ML in the hybrid
system can be obtained by the Dyson equation
\begin{subequations}
\begin{eqnarray}\label{dyson}
{\mathcal G}_{\text{TMD}}&=&G^0_{\text{TMD}}+G^{0}_{\text{TMD}}\Sigma\,
{\mathcal G}_{\text{TMD}} \quad , \\ \label{sigma}
\Sigma(\mathbf{k},i\omega_n) &=& \sum_{\mathbf q} T_{\mathbf{k},\mathbf{q}}
G^{0}_{\rm SC}({\mathbf q},i\omega_n) T^\dag_{\mathbf{k},\mathbf{q}} \quad .
\end{eqnarray}
\end{subequations}
By means of Eq.~(\ref{dyson}), the Green's function ${\mathcal G}_{\rm TMD}$ of the
TMD-ML in the presence of tunnel coupling is found as
\begin{eqnarray}
{\mathcal G}_{\mathrm{TMD}}({\mathbf k},i\omega_n)&=&\left[ (G^{0}_{\mathrm{TMD}}
({\mathbf k}, i\omega_n))^{-1} - \Sigma({\mathbf k},i\omega_n)\right]^{-1}.\quad
\end{eqnarray}
However, for a weakly coupled TMD-ML/S hybrid system, we have ${\mathcal 
G}_{\mathrm{TMD}}({\mathbf k},i\omega_n) \approx G^{0}_{\mathrm{TMD}}({\mathbf k},
i\omega_n) + \delta {\mathcal G}^{(2)}_{\mathrm{TMD}}(\mathbf{k},i\omega_n)$ to a good
approximation, where
\begin{equation}
\label{correction}
\delta {\cal G}^{(2)}_{\rm TMD}(\mathbf{k},i\omega_n)=G^{0}_{\mathrm{TMD}}(\mathbf{k},
i\omega_n) \Sigma(\mathbf{k},i\omega_n) G^{0}_{\rm TMD}(\mathbf{k},i\omega_n).  
\end{equation}
The superconducting correlations induced in the TMD are manifested by a non-vanishing
anomalous part of $\delta {\cal G}^{(2)}_{\mathrm{TMD}}$ given in
Eq.~(\ref{correction}). After some straightforward algebra, the intra-valley and
inter-valley components of the anomalous Green's functions for the TMD-ML can be cast
in the form
\begin{subequations}
\begin{eqnarray}
\label{fzzp}
&&\hat{\cal F}^{\zeta\zeta'}({\mathbf k},i\omega_n)=t_\zeta t_{\zeta'} 
\hat{g}_{e}^{(\zeta)}({\mathbf k},i\omega_n) \hat{f} (i\omega_n) 
\hat{g}_{h}^{(\zeta')}({\mathbf k},i\omega_n),\quad  \\
&&\hat{f} (i\omega_n) \equiv \sum_{\mathbf q}\hat{f} ({\mathbf q},i\omega_n)=
\begin{pmatrix}
0&\bar{f}_{\rm SC}(i\omega_n) \\
-\bar{f}_{\rm SC}(i\omega_n)&0
\end{pmatrix}. \quad 
\end{eqnarray}
\end{subequations}

\section{Results and Discussion}
\begin{table}[b]
  \begin{tabular}{| c | c | c | c |}
    \hline
    ~OP~ & spin & valley & $\omega$ \\ \hline
    $d^{(\text{e})}_0$ & ~singlet~ & symmetric & ~even~ \\ \hline  
    $d^{(\text{o})}_0$ & singlet & antisymmetric & odd \\ \hline
    ${\bf d}^{(\text{o})}$ & triplet & symmetric & odd \\ \hline
    ${\bf d}^{(\text{e})}$ & triplet & antisymmetric & even \\ \hline
    \end{tabular}
 \caption{\label{tab:symmetry}%
Classification of \textit{s}-wave OPs according to symmetries in their spin and valley 
 degrees of freedom, as well as their frequency dependence.}
\end{table} 

The retarded anomalous Green's function can be obtained from Eq.~(\ref{fzzp}) by analytic 
continuation, i.e. $i\omega_n \rightarrow \omega+i 0^+$. In the following, for the sake 
of brevity we will omit to write the small imaginary part. Similarly to the usual 
conventions applied to superconducting order parameters~\cite{mackenzie03}, the retarded 
anomalous Green's function that represents the proximity-induced superconducting 
correlations can be decomposed into singlet and triplet components as
\begin{eqnarray}
\hat{\cal{F}}^{\zeta\zeta'}({\mathbf k},\omega)&=&\left[ d_0^{\zeta\zeta'}({\mathbf 
k},\omega) \hat{\sigma}_0+{\mathbf d}^{\zeta\zeta'}({\mathbf k},\omega)\cdot\hat{\bm 
\sigma} \right] i \hat{\sigma}_y \quad , \nonumber \\
&\equiv&\begin{pmatrix}
-d_x+id_y& d_0+d_z\\
-d_0+d_z & d_x+i d_y,
\end{pmatrix} \quad . \label{f-vs-d}
\end{eqnarray} 
The singlet and triplet components of the superconducting correlations are parametrized
by the scalar $d_0$ and the vector ${\mathbf d}$, respectively. The components $d_x$
and $d_y$ describe equal-spin triplet pairing, while $d_0$ and $d_z$ correspond to
opposite-spin pairing. 
Since the overall wave function of the Cooper pairs must be antisymmetric due to the
fermionic nature of electrons, the superconducting order parameter (OP) can have
different symmetries with respect to each single-particle degree of freedom.
Equation~(\ref{fzzp}) is symmetric in $\mathbf{k}$, and the overall fermionic
anti-symmetry is realised as shown in Table~\ref{tab:symmetry} where the OPs are
classified according to their symmetries with respect to spin, valley, and frequency.
In what follows, we will show how various exotic triplet components can be induced in
a TMD-ML via proximity to an \textit{s}-wave singlet superconductor and in the presence
of an exchange splitting. In particular, \emph{intra-valley} and \emph{inter-valley}
pairings with both opposite- and equal-spin components emerge as a result of the
interplay of superconductivity with magnetic exchange and SOC. 

We can express the various component of the anomalous Green's function as 
\begin{widetext}
\begin{subequations}\label{eq:allGen}
\begin{align}
\label{eq:dogen}
d^{\zeta_1\zeta_2}_0({\mathbf k},\omega)&=t_{\zeta_1}t_{\zeta_2} \bar{f}_{\rm SC}
(\omega)\left[\mathbf{J}_{\text{eff}}^{\zeta_1}\cdot\mathbf{J}_{\text{eff}}^{\zeta_2}
+\omega^2-E_{\mathbf k}^2\right] \left[E^{\zeta_1\zeta_2}(\omega)+O^{\zeta_1\zeta_2}
(\omega)\right] \,\, , \\ \label{eq:dvecgen}
\mathbf{d}^{\zeta_1\zeta_2}({\mathbf k},\omega)&=-t_{\zeta_1}t_{\zeta_2}\bar{f}_{\rm 
SC}(\omega)\left[\omega\left(\mathbf{J}_{\text{eff}}^{\zeta_1}+
\mathbf{J}_{\text{eff}}^{\zeta_2}\right)+i \mathbf{J}_{\text{eff}}^{\zeta_1}\times 
\mathbf{J}_{\text{eff}}^{\zeta_2}+E_{\bf k}\left(\mathbf{J}_{\text{eff}}^{\zeta_1}-
\mathbf{J}_{\text{eff}}^{\zeta_2} \right) \right] \left[E^{\zeta_1\zeta_2}(\omega) 
+O^{\zeta_1\zeta_2}(\omega)\right]\,\, ,
\end{align}
\end{subequations}
\end{widetext}
where $\zeta_1$ and $\zeta_2$ are general valley indices, $E^{\zeta_1\zeta_2}(\omega)=
E^{\zeta_2\zeta_1}(\omega)$ is an even function of $\omega$, and $O^{\zeta_1\zeta_2}
(\omega)=-O^{\zeta_2\zeta_1}(\omega)$ is an odd function of $\omega$. The functions 
$E^{\zeta_1 \zeta_2}(\omega)$ and $O^{\zeta_1\zeta_2}(\omega)$ can be written as
\begin{subequations}
\begin{align}
E^{\zeta_1\zeta_2}(\omega)=\frac{1}{2}\left(\frac{1}{\Gamma^{\zeta_1 \zeta_2}(\omega)}+
\frac{1}{\Gamma^{\zeta_2 \zeta_1}(\omega)}\right) \quad , \\
O^{\zeta_1\zeta_2}(\omega)=\frac{1}{2}\left(\frac{1}{\Gamma^{\zeta_1 \zeta_2}(\omega)}-
\frac{1}{\Gamma^{\zeta_2 \zeta_1}(\omega)}\right) \quad ,
\end{align}
with 
\begin{align}\label{eq:Gamma}
\Gamma^{\zeta_1 \zeta_2}(\omega)= \left[\left( 
\mathbf{J}_{\text{eff}}^{\zeta_1}\right)^2 -\left(
\omega-E_{\mathbf{k}}\right)^2\right] \left[\left( 
\mathbf{J}_{\text{eff}}^{\zeta_2}\right)^2 -\left(
\omega+E_{\mathbf{k}}\right)^2\right].
\end{align}
\end{subequations}
The form of Eqs.~(\ref{eq:allGen}) enables straightforward identification of
superconducting correlations that are even or odd in frequency. We then define order
parameters for the various even-frequency superconducting correlations via their
$\omega\to 0$ limit; $\hat{\cal{F}}^{\zeta\zeta'}({\mathbf{k}},\omega=0)$. Similarly,
the order parameters of odd-frequency pairings are defined as the $\omega\to 0$ limit
of their derivative w.r.t.\ frequency; $\left. \partial_\omega \hat{\cal{F}}^{\zeta 
\zeta'}({\mathbf{k}},\omega)\right|_{\omega=0}$. In the following, we discuss the cases
of proximity-induced intra-valley ($\zeta_1 = \zeta_2$) and inter-valley ($\zeta_1 =
- \zeta_2$) pair correlations, separately.

\subsection{Intra-valley pairing} 

At first sight, intra-valley superconducting pairing may seem counter-intuitive because
it implies Cooper pairs having finite momentum. However, the existence of such pair
correlations is not forbidden by fundamental symmetries~\cite{fflo-rev,fradkin15}, and
instances of pair-density-wave order within individual valleys have been pointed out
recently for graphene~\cite{roy10,zhou13}, Weyl semimetals~\cite{cho12}, and a generic
multi-valley model system~\cite{tsu16}. Setting $\zeta_1=\zeta_2\equiv\zeta$ in Eqs.\
(\ref{eq:dogen}) and (\ref{eq:dvecgen}) yields the following results:
\begin{subequations}
\begin{eqnarray}
d^{\zeta\zeta}_0({\bf k},\omega)&=&t_\zeta^2 \bar{f}_{\rm SC}(\omega)\,
\frac{(\mathbf{J}_{\text{eff}}^\zeta)^2+\omega^2-E_{\bf k}^2}{\Gamma^{\zeta\zeta}
(\omega)} \quad , \label{eq:d0zz} \\
{\mathbf d}^{\zeta\zeta}({\mathbf k},\omega)&=&t_\zeta^2\bar{f}_{\rm SC}(\omega)\,
\frac{-2\omega \mathbf{J}_{\text{eff}}^\zeta }{\Gamma^{\zeta\zeta}(\omega)} \quad ,
\label{eq:dveczz}
\end{eqnarray}
\end{subequations}
where we have used $E^{\zeta\zeta}(\omega)=1/\Gamma^{\zeta\zeta}(\omega)$ and
$O^{\zeta\zeta}(\omega)=0$. Clearly, the singlet and triplet components for
intra-valley pairing are even- and odd-frequency, respectively. This is consistent with
the fact that the even symmetry of the components $d_{x,y,x}^{\zeta\zeta}$ with respect
to the valley degree of freedom requires odd symmetry with respect to the frequency
according to Table~\ref{tab:symmetry}. Note, however, that the intra-valley pairings
from the $\mathbf{K}$ and $\mathbf{K'}$ points are generally different because of the
valley dependence of the effective spin splitting parametrized by
$\mathbf{J}_{\text{eff}}^\zeta$, as illustrated in Fig.~\ref{fig1}(c), and potentially
also because of a valley-dependent tunnel coupling $t_{\zeta}$. The proportionality of
the triplet pairing vector ${\mathbf d}^{\zeta\zeta}$ to the effective spin splitting
${\bf J}^{\zeta}_{\rm eff}$ in each valley renders this exotic order parameter highly
tunable. In particular, $d_{x,y}^{\zeta\zeta}$ will only be finite in the presence of
finite external exchange-field components $J_{x,y}$. In contrast, SOC results in a
finite $d_z^{\zeta\zeta}$ even in the absence of any exchange field.

\begin{figure}[b]
\includegraphics[width=0.95\columnwidth]{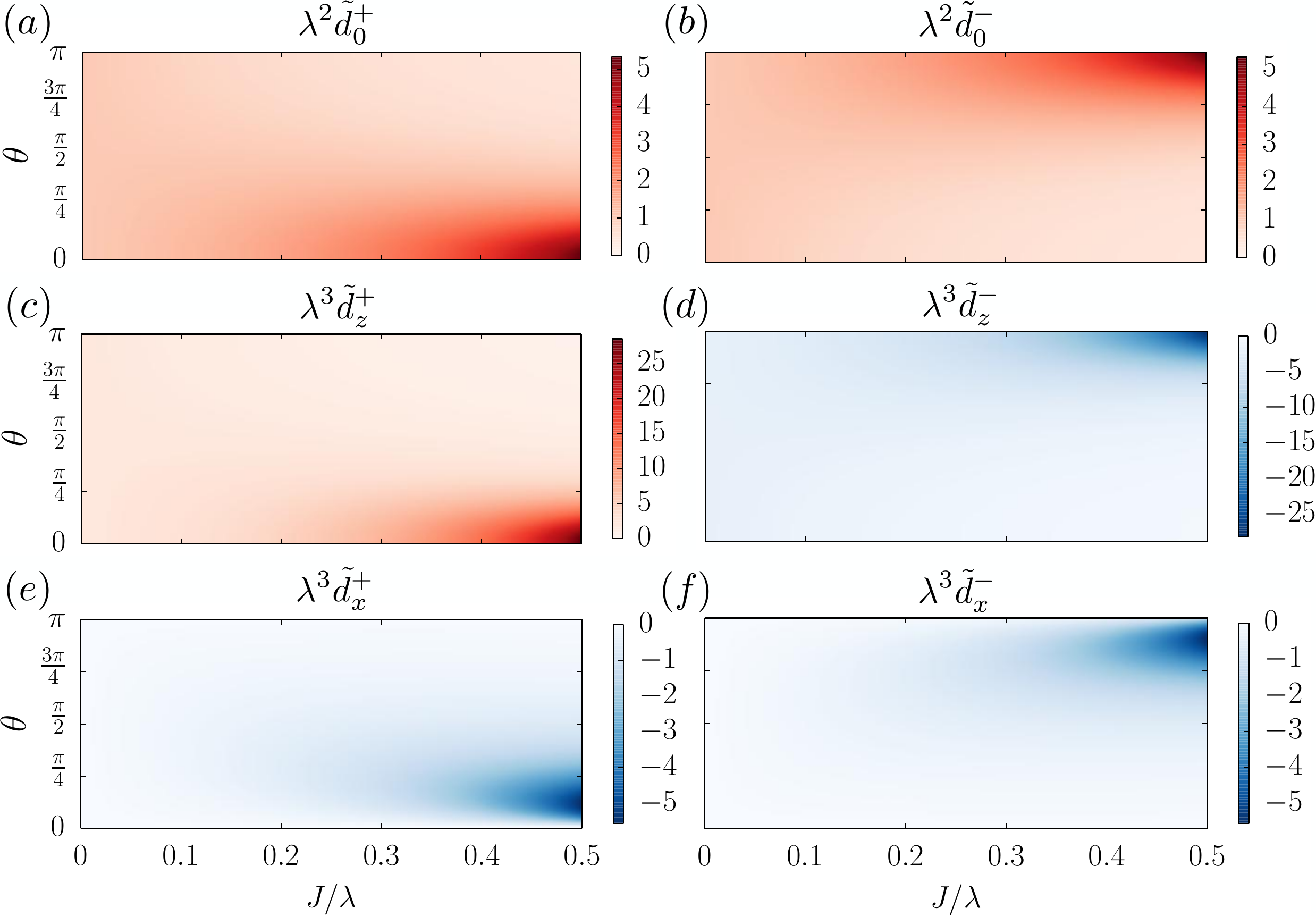}
\caption{\label{fig2}%
Magnitude of intra-valley superconducting order parameters from Eqs.~(\ref{eq:intraOPs})
induced in a monolayer TMD when the external exchange field is $\mathbf{J} = (J_x, 0,
J_z)$. Here $J=\sqrt{J_x^2+J_z^2}$, $\theta=\arctan(J_x/J_z)$, and $E_\mathbf{k} =
-0.25\, \lambda$.}
\end{figure}

To illustrate the parametric dependences of the intra-valley order parameters, we 
consider the quantities,
\begin{subequations}\label{eq:intraOPs}
\begin{eqnarray}
\tilde d_0^\zeta &=& \frac{d^{\zeta\zeta}_0({\mathbf k},0)}{t_\zeta^2
\bar{f}_{\rm SC}(0)}\equiv \left[\left( \mathbf{J}_{\text{eff}}^{\zeta}\right)^2 -
E^2_{\mathbf{k}}\right]^{-1} \,\, , \\
{\tilde{\mathbf d}}^\zeta &=& \frac{\left. \partial_\omega\, {\mathbf d}^{\zeta\zeta}
({\mathbf k},\omega)\right|_{\omega=0}}{t_\zeta^2\bar{f}_{\rm SC}(0)}
\equiv \frac{-2 \mathbf{J}_{\text{eff}}^\zeta}{\left[\left( 
\mathbf{J}_{\text{eff}}^{\zeta}\right)^2 - E^2_{\mathbf{k}}\right]^2}\,\, . \quad
\end{eqnarray}
\end{subequations}
Both $\tilde d_0^\zeta$ and ${\tilde{\mathbf d}}^\zeta$ are representative because
the frequency dependence of superconducting correlations near the chemical potential is
weak. Figure~\ref{fig2} shows results obtained for the situation when $\mathbf{J}$ is
in the $xz$ plane. For definiteness, we choose $\mathbf{k}$ such that $E_\mathbf{k} =
-0.25\,\lambda$, but the same qualitative behavior is exhibited for other values.
Note that, to be able to plot dimensionless quantities, the effective order parameters
that are even and odd in frequency have been scaled with $\lambda^2$ and $\lambda^3$,
respectively. 

The singlet pairings $\tilde d_0^\zeta$ are finite for any magnitude $J$ of the exchange 
field and direction parametrized by the angle $\theta=\arctan(J_x/J_z)$. However, the
existence of triplet components $\tilde d_z^\zeta$ and $\tilde d_x^\zeta$, which are 
between opposite and equal spins, requires a finite exchange splitting ${\bf J}$ with
out-of-plane and in-plane components, respectively. For large enough values of $J$,
rapid increases in magnitude occur for all OPs whenever a divergence due to vanishing
denominators in Eqs.~(\ref{eq:intraOPs}) is approached.~\footnote{In real systems,
such divergences are cut off by the finite quasiparticle lifetimes.} On the other hand, 
by changing the orientation of the external exchange field, the opposite-spin components 
at the two valleys reach maximum strengths for $\pm z$ alignments of ${\bf J}$,
respectively. The strength of equal-spin pairings $d_x^{\zeta}$ can also be varied by 
$\theta$, and the maximum strengths for the different valleys $K$ and $K'$ are attained 
for close-to-opposite out-of-plane alignments $\theta\gtrsim 0$ and $\theta \lesssim 
\pi$, respectively. Therefore, careful adjustment of the exchange field makes it 
possible to engineer situations where particular OP types dominate, rendering the hybrid 
system under consideration a versatile laboratory for studying unconventional 
superconductivity. Furthermore, the spin part of the triplet-Cooper-pair wave function 
$\ket{\psi^\zeta_\mathrm{t}}$ for each valley is fully determined by the vector 
$\mathbf{d}^{\zeta\zeta}$ as described, e.g., in Ref.~\onlinecite{mackenzie03}. For the 
system under study the triplet wave functions can therefore be manipulated directly via 
the effective exchange fields as
\begin{eqnarray}
\ket{\psi^\zeta_\mathrm{t}} &\propto& J^\zeta_{\mathrm{eff}, x} (-\ket{\uparrow
\uparrow} + \ket{\downarrow\downarrow}) \nonumber \\[0.1cm] && \hspace{0.3cm} +
i J^\zeta_{\mathrm{eff}, y} (\ket{\uparrow\uparrow} + \ket{\downarrow\downarrow}) +
J^\zeta_{\mathrm{eff}, z} (\ket{\uparrow\downarrow} + \ket{\downarrow\uparrow}) \, .
\quad 
\end{eqnarray}

\subsection{Inter-valley pairing}

For the case $\zeta_1=-\zeta_2\equiv\zeta$, the proximity-induced superconducting
pairings parametrized by Eqs. (\ref{eq:dogen}) and (\ref{eq:dvecgen}) exhibit a very
rich behavior that can also be manipulated by the external exchange field ${\mathbf 
J}$. In particular, each order parameter has both even and odd-frequency components in
general. The symmetries with respect to the spin and valley degrees of freedom,
together with the even- or odd-frequency nature of the corresponding OPs, is consistent
with the classifications given in Table~\ref{tab:symmetry}. Of particular interest is
the emergence of the even-frequency equal-spin triplet component $\propto
\mathbf{J}_{\text{eff}}^{\zeta}\times\mathbf{J}_{\text{eff}}^{-\zeta}\equiv 2\zeta 
\lambda\, \mathbf{J}\times\mathbf{\hat z}$, which arises purely because of the
noncollinearity of the effective spin-splitting fields in the two valleys and is
therefore a direct consequence of the interplay between SOC and the external exchange
field. Furthermore, an odd-frequency singlet OP emerges when the effective exchange
fields for the two valleys are different. This is analogous to the situation
encountered previously in double quantum-dot systems in contact with an
\textit{s}-wave superconductor~\cite{sothmann14,tanaka16}.

To illustrate again the parametric dependencies of the various OPs, we define the 
following quantities,
\begin{widetext}
\begin{subequations}
\begin{align}
%%%%%%
\tilde d_0^{(\mathrm{e})} &\equiv \frac{d^{\zeta,-\zeta}_0({\mathbf k},
0)}{t_+ t_- \bar{f}_{\rm SC}(0)}=  \frac{\mathbf{J}_{\text{eff}}^{+}\cdot
\mathbf{J}_{\text{eff}}^{-}-E_{\mathbf k}^2}{\Gamma^{+-}(0)}
 \quad ,\\
%%%%%%%
\tilde d_0^{(\mathrm{o})\zeta} &\equiv \frac{\left. \partial_\omega\, d_0^{\zeta,-\zeta}
({\mathbf k},\omega)\right|_{\omega=0}}{t_+ t_- \bar{f}_{\rm SC}(0)} =  \zeta 
\left(\mathbf{J}_{\text{eff}}^{+}\cdot\mathbf{J}_{\text{eff}}^{-}-E_{\mathbf k}^2
\right) \frac{2 E_{\mathbf k} \left[ \big( \mathbf{J}_{\text{eff}}^{+}\big)^2 - 
\big(\mathbf{J}_{\text{eff}}^{-}\big)^2\right]}
{\left[\Gamma^{+-}(0)\right]^2}
 \quad , \\
%%%%%%%%%
{\tilde{\mathbf d}}^{(\mathrm{e})\zeta} &\equiv \frac{{\mathbf d}^{\zeta,
-\zeta}({\mathbf k},0)}{t_+ t_- \bar{f}_{\rm SC}(0)}= -\zeta\frac{i 
\mathbf{J}_{\text{eff}}^{+}\times \mathbf{J}_{\text{eff}}^{-}-2 \lambda E_{\bf k} 
\hat{\mathbf{z}}}
{\Gamma^{+-}(0)}
\quad , \\
%%%%%%%%%
{\tilde{\mathbf d}}^{(\mathrm{o})} & \equiv
\frac{\left. \partial_\omega\, \mathbf{d}^{\zeta,
-\zeta}({\mathbf k},\omega)\right|_{\omega=0}}{t_+ t_- \bar{f}_{\rm SC}(0)}= 
\frac{-2\mathbf{J}}{\Gamma^{+-}(0)}
-\left[i\mathbf{J}_{\text{eff}}^{+}\times \mathbf{J}_{\text{eff}}^{-}- 2 \lambda
E_{\bf k}\hat{\mathbf{z}}\right]\frac{2 E_{{\mathbf k}} \left[ \big( 
\mathbf{J}_{\text{eff}}^{+}\big)^2 - \big(\mathbf{J}_{\text{eff}}^{-}\big)^2
\right]}{\left[ \Gamma^{+-}(0) \right]^2}\quad ,
\end{align}
\end{subequations}
\end{widetext}
with $\Gamma^{+-}(0)$ defined by Eq. (\ref{eq:Gamma}). As is apparent from the above 
relations, the odd-frequency singlet OP and the even-frequency triplet component are 
antisymmetric with respect to the exchange of valleys and, consequently, have no 
corresponding intra-valley OP realization. In order to illustrate more clearly the
structure of valley-odd inter-valley OPs, we write down the explicit form of the
Cooper-pair wave function for the spin-triplet even-frequency component:
\begin{eqnarray}
\ket{\psi^{(e)}_\mathrm{t}} &\propto&
\lambda (\ket{+-}-\ket{-+})\otimes 
\scalebox{1.4}{$[$}\, i J_{x}
(-\ket{\uparrow
\uparrow} + \ket{\downarrow\downarrow}) \nonumber \\[0.1cm] && \hspace{0.3cm} +
J_{y} (\ket{\uparrow\uparrow} + \ket{\downarrow\downarrow}) +
E_k (\ket{\uparrow\downarrow} + \ket{\downarrow\uparrow})\,\scalebox{1.4}{$]$} \, .
\quad 
\end{eqnarray}
As expected from symmetry arguments given above, this wave function is a valley-singlet
$\propto (\ket{+-}-\ket{-+})$ and vanishes in the absence of SOC, as the latter is
responsible for the valley-symmetry breaking.

\begin{figure}[t]
\includegraphics[width=0.95\columnwidth]{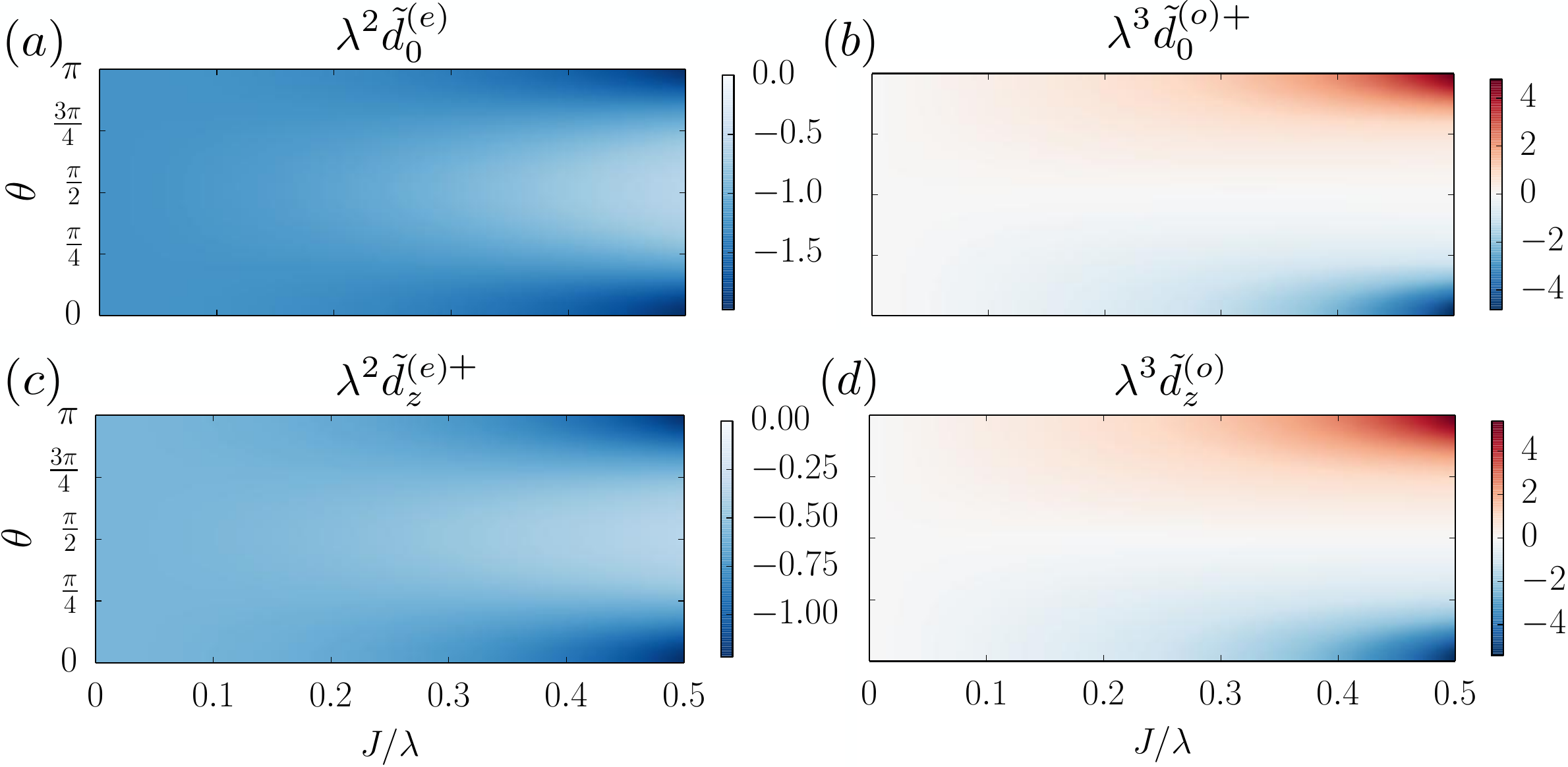}
\caption{\label{fig3}%
Magnitude of inter-valley opposite-spin superconducting order parameters induced in
a monolayer TMD when the external exchange field is $\mathbf{J} = (J_x, 0, J_z)$.
Shown are dependences on $J=\sqrt{J_x^2+J_z^2}$ and $\theta=\arctan(J_x/J_z)$, keeping
$E_\mathbf{k} = -0.25\, \lambda$ fixed. }
\end{figure}

\begin{figure}[b]
\includegraphics[width=0.7\linewidth]{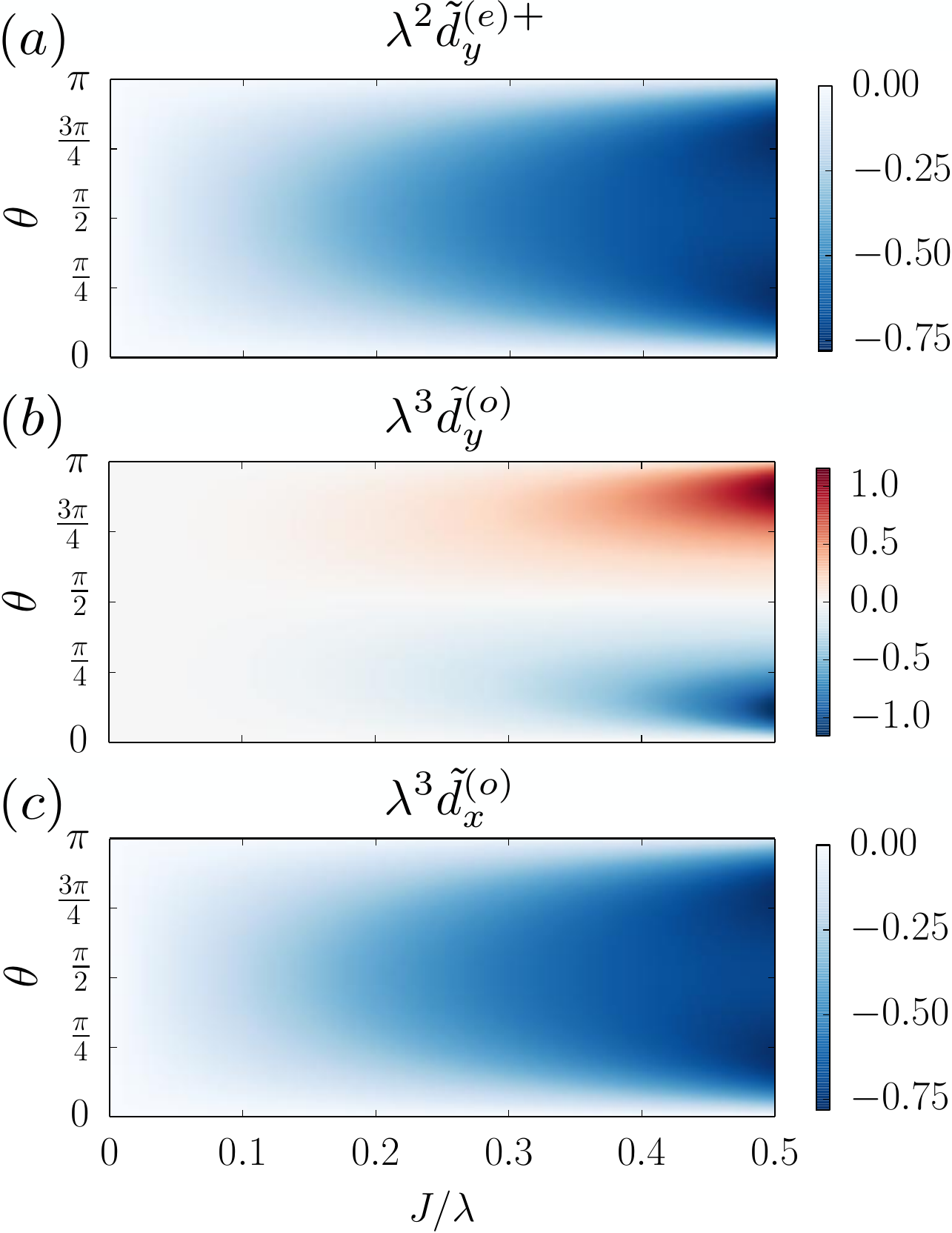}
\caption{\label{fig4}%
Magnitude of inter-valley equal-spin superconducting order parameters for the same
situation and conventions as in Fig.~\ref{fig3}.}
\end{figure}

Figures~\ref{fig3} and \ref{fig4} show the dependences of opposite-spin and equal-spin
OPs, respectively, on the magnitude and direction of an external exchange field that
is in the $xz$ plane. Again $E_\mathbf{k} = -0.25\,\lambda$ was assumed when
calculating these plots. While the even-frequency components $\tilde{d}_{0}^{(e)}$
and $\tilde{d}_{z}^{(e)}$ shown in Figs.~\ref{fig3}(a),(c) are finite even for small 
${\bf J}$, the corresponding odd-frequency OPs that can be seen in
Figs.~\ref{fig3}(b),(d) require the presence of an external exchange field. Moreover,
all opposite-spin pairings have their largest magnitude for ${\bf J}$ perpendicular to 
the plane of the TMD-ML, i.e., when $\theta\sim 0$ and $\pi$.

Inspection of Fig.~\ref{fig4}(a) reveals  that the even-frequency equal-spin triplet OP 
represented by ${\tilde d}^{(\mathrm{e})\zeta}_y$ has completely different behavior 
compared to opposite-spin terms shown in Fig. \ref{fig3}. In particular it becomes 
dominant in a broad range around $\theta \approx \pi/2$ when $J/\lambda \gtrsim 0.2$,
for the chosen parameter regime. The odd-frequency equal-spin OP ${\tilde 
d}^{(\mathrm{o})}_x$ shown in Fig.~\ref{fig4}(c) exhibits the same parametric dependences 
as ${\tilde d}^{(\mathrm{e})\zeta}_y$, whereas ${\tilde d}^{(\mathrm{o})}_y$ 
[Fig.~\ref{fig4}(b)] emerges only for large-enough $J/\lambda$ and for certain angles 
$\theta$ close to (but not equal to) $0$ or $\pi$. Generally, the equal-spin OPs are
only finite when $\mathbf{J}$ has a finite in-plane component. The plethora of OPs
realized for inter-valley pairing is a direct consequence of the, in general,
noncollinear effective exchange fields $\mathbf{J}_\mathrm{eff}^\pm$ acting in the two
valleys, which can again be tailored to generate specific superconducting pairing types.

\subsection{Discussion}

We have identified a great variety of superconducting OPs arising from the coupling
of the TMD-ML to both magnetic and superconducting materials. Two basic ingredients
are crucial for facilitating the unconventional types of pairing. First, the
existence of two valleys enables symmetric and antisymmetric inter-valley pairing
mechanisms so that even- and odd-frequency behavior becomes, in principle, possible
for any type of superconducting OP. Secondly, the interplay of SOC and exchange
splitting enables triplet pairings to emerge. Thus our system of interest differs
markedly from previously considered superconductor-ferromagnet hybrid systems where
a spatially nonuniform magnetization or a Rashba-type SOC gave rise to equal-spin
triplet pairing. Our work also extends recent studies of intrinsic~\cite{yuan14,law16}
and proximity-induced~\cite{rossi16} superconducting phases in a TMD-ML.

We find that the components of the exotic even-frequency triplet OP ${\tilde{\mathbf 
d}}^{(\mathrm{e})\zeta}$ can be comparable in magnitude to the conventional
even-frequency spin-singlet OPs for intra- and inter-valley pairing. Separating 
${\tilde{\mathbf d}}^{(\mathrm{e})\zeta} = {\tilde{\mathbf d}}^{(\mathrm{e}) 
\zeta}_\parallel + {\tilde{\mathbf d}}^{(\mathrm{e})\zeta}_\perp$ in terms of the
equal-spin and opposite-spin contributions
\begin{subequations}
\begin{eqnarray}
{\tilde{\mathbf d}}^{(\mathrm{e})\zeta}_\parallel &=& \frac{-2i \zeta \lambda\, 
\mathbf{J}\times\mathbf{\hat z}}{\left[\left(\mathbf{J}_{\text{eff}}^{+}\right)^2 - 
E^2_{\mathbf{k}}\right]\left[\left(\mathbf{J}_{\text{eff}}^{-}\right)^2 - 
E^2_{\mathbf{k}} \right]} \quad , \\
{\tilde{\mathbf d}}^{(\mathrm{e})\zeta}_\perp &=& \frac{2\zeta\lambda\, E_\mathbf{k}\,
\mathbf{\hat z}}{\left[\left(\mathbf{J}_{\text{eff}}^{+}\right)^2 - E^2_{\mathbf{k}}
\right]\left[\left(\mathbf{J}_{\text{eff}}^{-}\right)^2 - E^2_{\mathbf{k}} \right]}
\quad ,
\end{eqnarray}
\end{subequations}
the part ${\tilde{\mathbf d}}^{(\mathrm{e})\zeta}_\perp$ can be identified with the
OP previously associated with intrinsic Ising superconductivity~\cite{law16}
The equal-spin part ${\tilde{\mathbf d}}^{(\mathrm{e})\zeta}_\parallel$ has the same 
antisymmetric behavior with respect to the exchange of valleys, and its direct
dependence on the exchange field $\mathbf J$ makes it highly tunable. Extending recent 
experimental studies~\cite{ising-1,ising-3,ising-2} of Ising superconductivity to 
situations where the TMD-ML is coupled to a magnetic insulator would facilitate 
exploration of this new OP. The prefactor $\lambda$ clearly indicates that finite SOC
is an essential ingredient for both opposite-spin and equal-spin contributions to 
${\tilde{\mathbf d}}^{(\mathrm{e})\zeta}$.

It is worth noting that, generally, strong dependences on the valley index are 
exhibited by intra-valley OPs. This behavior results from the different effective 
exchange fields being present in each valley determining the pairing magnitudes. The 
valley-dependent proximity-induced superconductivity in the presence of a finite
exchange field is a new manifestation of coupled spin-valley physics of TMDs. Thus
complementing previous proposals for spintronics applications of TMD-MLs, our results 
indicate that hybrid systems made of magnetic TMD-MLs and conventional superconductors 
can be very promising systems for further investigations within the field of
\emph{super-spintronics}~\cite{linder15}.

Finally, we would like to comment on possible experimental observations of the effects
found here. First of all, the setup we have proposed as a TMD-ML tunnel-coupled to a
superconducting layer is realizable using current experimental techniques. Then, in
order to recognize the various OPs having different symmetries with respect to spin,
valley index and frequency, one should be able to identify the distinctive signatures
of each component using, e.g., tunneling and Andreev spectroscopies, as these methods
measure the local density of states and consequently can reveal the amplitude of the
induced superconducting gap \cite{tinkham04,wolf12}. For example, according to our
results, one should observe an enhancement of the superconducting gap as the $z$
component of the external exchange splitting is increased, in contrast to the case of
conventional superconductivity. On the other hand, the existence of odd-frequency OPs
and their respective variations can also be observed using tunneling spectroscopy
\cite{asano07,robinson15} via the energy dependence of odd-frequency pairings and their
tunability with the external exchange field. Furthermore, scanning Josephson
spectroscopy~\cite{Smakov01,Randeria16} can be employed to access the phase properties
of the superconducting state. Lastly, spin-polarized scanning tunneling 
spectroscopy~\cite{Wiesendanger09} can reveal the presence of spin-triplet pairings
through the suppression of Andreev reflection in the opposite-spin channel
\cite{simon16}. 

\section{Conclusions}

We have investigated the proximity-induced superconductivity in monolayers of
transition metal dichalcogenides that are tunnel-coupled to a conventional singlet 
\textit{s}-wave superconductor and subject to a magnetization-induced spin splitting. 
Various order parameters have been identified, exhausting all possible symmetries with
respect to spin, valley, and frequency dependence. In particular, even-frequency
inter-valley triplet pairing is facilitated by the intrinsic spin-orbit coupling in
the transition metal dichalcogenide material. The opposite-spin-pairing component of
this exotic superconducting correlation is a realization of the previously discussed
phenomenon of Ising superconductivity, while the equal-spin components are only finite
when the exchange field has in-plane components.

We obtained analytical results that clearly illustrate the dependence of all possible
superconducting order parameters on relevant parameters such as the magnitude and
direction of the exchange field as well as the strength of the intrinsic spin-orbit
coupling. Tailoring these parameters is shown to provide great selectivity to access
dominant superconducting correlations in different regimes. We hope that this detailed
insight into the behavior of an experimentally accessible system will facilitate
further systematic exploration of exotic superconductivity in hybrid structures.

\acknowledgments
A.~G.~M.\ acknowledges financial support from the Iran Science Elites
Federation under Grant No.\ 11/66332.

\bibliography{scTMD.bib}

\end{document}